\documentclass[aps,preprint]{revtex4}%
\usepackage{amssymb}
\usepackage{amsfonts}%
\usepackage{amsmath}%
\usepackage{graphicx}
\providecommand{\U}[1]{\protect\rule{.1in}{.1in}}
\begin{document}
\title{Pressure-induced superconductivity in europium metal}
\author{M. Debessai, T. Matsuoka$^{\ast}$, J. J. Hamlin$^{\ast\ast}$, and J. S. Schilling}
\affiliation{Department of Physics, Washington University, CB 1105, One Brookings Drive,
St. Louis, Missouri 63130, USA}
\author{K. Shimizu}
\affiliation{KYOKUGEN, Center for Quantum Science and Technology under Extreme Conditions,
Osaka University, 1-3 Machikaneyama, Toyonaka, Osaka 560-8531, Japan}

\begin{abstract}
Divalent Eu (4f$^{7}$, $J=7/2$) possesses a strong local magnetic moment which
suppresses superconductivity. Under sufficient pressure it is anticipated that
Eu will become trivalent (4f$^{6}$, $J=0$) and a weak Van Vleck paramagnet,
thus opening the door for a possible superconducting state, in analogy with Am
metal (5f$^{6}$, $J=0$) which superconducts at 0.79 K. We present ac
susceptibility and electrical resistivity measurements on Eu metal for
temperatures 1.5 - 297 K to pressures as high as 142 GPa. At approximately 80
GPa \ Eu becomes superconducting at $T_{c}\simeq$ 1.8 K; $T_{c}$ increases
linearly with pressure to 2.75 K at 142 GPa. Eu metal thus becomes the
53$^{rd}$ known elemental superconductor in the periodic table.$\vspace
{1.2cm}\vspace{1.2cm}\vspace{1cm}\vspace{1cm}$\newline$^{\ast}$Permanent
address: \ KYOKUGEN, Center for Quantum Science and Technology under Extreme
Conditions, Osaka University, 1-3 Machikaneyama, Toyonaka, Osaka 560-8531,
Japan\vspace{0.2cm}\newline$^{\ast\ast}$Present address: \ Department of
Physics and Institute for Pure and Applied Physical Sciences, University of
California, San Diego, La Jolla, CA 92093

\end{abstract}
\date{January 12, 2009}
\maketitle

Bernd Matthias once mused that all nonmagnetic metals might become
superconducting, if only they be cooled to a sufficiently low temperature
\cite{matthias}. Of the 92 naturally occurring elements in the periodic table,
there are 30 known elemental superconductors at ambient pressure and 22 more
that become superconducting under high pressure \cite{periodic table}. An
intriguing question is whether the remaining 40 elemental solids become
superconducting in some temperature/pressure range. One could, in fact, pose
this same question for all solids.

Across the entire lanthanide series, only its first member, La, superconducts
at ambient pressure. The reasons for this appear to be twofold: \ (1) the
local-moment magnetism in all lanthanides except La, Yb, and Lu leads to
strong pair-breaking effects, and (2) as for non-superconducting Sc or Y, the
relatively weak $d$-character of the conduction electrons for heavy
lanthanides like Lu results in an only diminutive pairing interaction. Since
compressing the lattice enhances the $d $-electron concentration, it is not
surprising that Sc, Y, and Lu all become superconducting under pressure
\cite{DebessaiPRB}; indeed, the vast majority of transition metals superconduct.

In contrast, the pressure-induced superconductivity observed for Ce metal
above 2 GPa arises from the suppression of its magnetism \cite{wittig}. The
fact that all lanthanides other than La, Ce, and Lu do not superconduct under
pressure, in spite of their enhanced $d$-electron concentration, is a tribute
to the stability of their strong local-moment magnetism. At sufficiently high
pressures, however, one would anticipate that the lanthanide valence should
increase as electrons are successively squeezed out of the 4f shell into the
$s,p,d$-conduction band. The first two lanthanide metals to do this would
likely be Eu and Yb since both are divalent at ambient pressure, in contrast
to all others which are trivalent.

Whereas trivalent Yb would exhibit a strong local magnetic moment by virtue of
its 4f$^{13}$ configuration with $J=L+S=3+%
%TCIMACRO{\U{bd}}%
%BeginExpansion
\frac12
%EndExpansion
=7/2,$ trivalent Eu would be left with a 4f$^{6}$ electron shell where $S=L=3$
and thus $J=L-S=0.$ Under sufficient pressure, therefore, the divalent
antiferromagnet Eu would be expected to become a trivalent weak Van Vleck
paramagnet and a good candidate for superconductivity \cite{rosengren},
perhaps at temperatures as high as 10 - 15 K as for the other trivalent
rare-earth superconductors La and Lu \cite{DebessaiPRB}. Support for this
possibility was given many years ago by Matthias \textit{et al.}
\cite{matthias1} who pointed out that the Laves-phase compound EuIr$_{2},$
where Eu is believed to be trivalent, does indeed superconduct below 3 K at
ambient pressure, as do the analogous nonmagnetic trivalent-ion systems
ScIr$_{2}$, YIr$_{2}$, LaIr$_{2}$, and LuIr$_{2}$. We note that the only other
known elemental metal with Van Vleck paramagnetism, trivalent Am with a
5f$^{6}$ electron shell, does indeed superconduct at $T_{c}\simeq$ 0.79 K at
ambient pressure, $T_{c}$ rising to 2.2 K at 6 GPa \cite{griveau}.

Estimates of the pressure necessary for the full divalent-to-trivalent
transition in Eu vary from 35 GPa by Rosengren and Johansson \cite{rosengren}
to 71 GPa by Min \textit{et al}. \cite{min}. X-ray diffraction studies to 30
GPa at room temperature reveal a bcc to hcp transition in Eu near 12.5 GPa
with a new closed-packed Eu-III phase appearing above 18 GPa \cite{takemura}.
Room temperature M\"{o}ssbauer-effect \cite{farrell,wortmann} and
L$_{\text{III}}$ absorption edge \cite{rohler} studies indicate that Eu's
valence $\nu$ increases rapidly with pressure from 2.0 to nearly 2.5 at 12
GPa; however, the latter measurements reveal that $\nu$ saturates at higher
pressures, reaching only $\nu\approx2.65$ at 34 GPa. Significantly higher
pressures are apparently necessary to bring Eu into its fully trivalent state.
In 1981 Bundy and Dunn \cite{Bundy} searched for a superconducting transition
in electrical resistivity measurements on Eu metal to pressures as high as 40
GPa; unfortunately, no superconductivity was observed above 2.3 K, the lowest
temperature of their measurement.

Using a diamond-anvil cell (DAC) we have carried out electrical resistivity
and ac susceptibility studies above 1.5 K on pure Eu metal to pressures as
high as 142 GPa. Above 70 - 80 GPa a superconducting transition appears near
1.7 K which increases slowly with pressure. Eu thus becomes the 53$^{rd}$
known elemental superconductor.

In the present electrical resistivity and ac susceptibility experiments, a
membrane-driven DAC \cite{JimDAC} was used with 1/6-carat, type Ia diamond
anvils with 0.18 mm culets beveled at 7$^{\circ}$ out to 0.35 mm. Disc-shaped
metal gaskets 0.25 mm thick and 3 mm diameter made of Re or BeCu-alloy were
chosen for the resistivity or ac susceptibility measurements, respectively.
The gaskets were preindented to 25 - 30 $\mu$m and a 90 $\mu$m diameter hole
was electro-spark drilled through the center of the gasket. The high-purity Eu
sample (99.98 \% metals basis), obtained from the Materials Preparation Center
of the Ames Laboratory \cite{ames}, was packed into the gasket hole\ together
with several tiny ruby spheres \cite{chervin} to allow the determination of
the pressure \textit{in situ} at 1.6 K from the $R_{1}$ ruby fluorescence line
with resolution 0.2 GPa using the revised pressure scale of of Chijioke
\textit{et al} \cite{chijioke}. The three highest pressures attained in the
present experiment (127, 135, and 142 GPa) were determined from the shift in
the diamond vibron \cite{akahama2} at the sample center since the ruby
fluorescence could no longer be resolved. No pressure medium was used.
Resistivity and susceptibility data obtained while warming were preferred
since they contain less noise than cooling data. For measurements to ambient
temperature the warming rate was typically 1-2 K/min, whereas in
determinations of the superconducting transition temperature the rate was
slowed to $\sim$ 1 K/h. Further details of the DAC techniques used in the
electrical resistivity \cite{shimizu1} and ac susceptibility
\cite{DebessaiPRB,JimDAC,matsuoka1} measurements are given elsewhere.

The electrical resistance of Eu at 297 K was found to increase monotonically
with pressure from 0.7 $\Omega$ at 10 GPa to 8 $\Omega$ at 27 GPa. As the
pressure was increased further, a Pt lead inside the cell failed, forcing us
to use an adjacent Pt lead as a combined current/voltage probe (quasi
four-point measurement); we estimate the contribution from this short section
of Pt lead to be only 0.1 $\Omega$; however, the contact resistance between Pt
lead and Eu sample may be much larger. As seen in Fig.~1, the resistance
continues to increase significantly with pressure to the highest pressure
reached, 91 GPa, not solely at 297 K but over the entire experimental
temperature range down to 1.5 K. That this resistance increase is at least
partially intrinsic, and not merely due to changes in sample dimension, defect
concentration, or contact resistance, is evidenced by the fact that R(T)
\textit{decreases} appreciably as the pressure is reduced from 91 to 62 GPa.
An increase in the resistivity of Eu with pressure was observed in earlier
studies to pressures as high as 40 GPa \cite{Stager,McWhan,Shelton,Bundy}.
Whether or not the bend in R(T) near 100 K in Fig.~1 is indicative of a
magnetic or structural phase transition can only be given a clear answer
through future temperature-dependent x-ray diffraction or M\"{o}ssbauer effect
studies to extreme pressures.

Particularly interesting are the data in Fig.~1 at 73, 81, and 91 GPa where a
sharp decrease in the resistance is seen upon cooling below 2 K (the inset
shows data on an expanded scale), hinting at a superconducting transition
which increases slowly with pressure. That the resistance does not fall to 0
$\Omega$ below the superconducting transition is not uncommon \cite{shimizu1}
and may arise from the Pt lead and its contact resistance to the Eu sample, as
well as from possible microcracks in the strongly plastically deformed sample
through uniaxial stresses. However, it is well known that the electrical
resistivity is a sensitive technique for detecting even trace concentrations
of a superconducting phase, but is poorly suited to establish whether or not a
material is a bulk superconductor. To this end the magnetic susceptibility is
a far superior diagnostic tool.

In Fig.~2 it is seen that at 76 GPa pressure the ac susceptibility shows no
evidence for superconductivity down to nearly 1.5 K. However, at 84 GPa a
sharp drop $\Delta\chi^{\prime}$ in the ac susceptibility\ appears at 1.78 K
(transition midpoint) which shifts slowly to higher temperatures with
increasing pressure, reaching 2.75 K at the highest pressure measured (142
GPa) \cite{note2}. Using the analysis discussed in detail in an earlier
publication \cite{DebessaiPRB}, the observed $\Delta\chi^{\prime}\approx$ 20
nV jump at $T_{c}$ is consistent with perfect diamagnetism, the hallmark of a
superconductor; in fact, in previous experiments on superconducting Y, Sc, and
Lu samples under nearly identical conditions we also find $\Delta\chi^{\prime
}\approx$ 15 - 20 nV \cite{DebessaiPRB}. In one experiment to pressures as
high as 94 GPa, the Eu sample became strongly oxidized through inadvertent
exposure to air; no diamagnetic transition was observed above 1.5 K, thus
confirming that the diamagnetic jump $\Delta\chi^{\prime}(T)$ seen in Fig.~2
does, in fact, originate from the Eu sample and not, for example, from the
CuBe gasket \cite{note1}.

In Fig.~3 the values of $T_{c}$ from the ac susceptibility (transition
midpoint) and electrical resistivity (low-temperature onset) are plotted
versus pressure; from the former studies $T_{c}$ is seen to increase linearly
with pressure at the very moderate rate of +18 mK/GPa. This should be compared
to the value of +360 mK/GPa for metallic Y \cite{HamlinY} or +900 mK/GPa for
pure Li \cite{shanti}.

AC susceptibility and electrical resistivity measurements were also carried
out under an applied dc magnetic field of 500 Oe. Within the experimental
resolution of 30 mK, no shift in the superconducting transition could be
resolved, implying that $\left\vert dT_{c}/dH\right\vert \leq0.06$ mK/Oe or
$\left\vert T_{c}^{-1}dT_{c}/dH\right\vert \leq3\times10^{-5}$ Oe$^{-1}$. This
upper limit is comparable to the relative shift in $T_{c}$ measured in
experiment for the actinide metal Am ($5\times10^{-5}$ Oe$^{-1}$)
\cite{griveau} as well as for the $d$-electron metals Sc ($1.6\times10^{-5}$
Oe$^{-1}$), Y ($5\times10^{-5}$ Oe$^{-1}$), and Lu ($6\times10^{-5}$ Oe$^{-1}%
$) \cite{DebessaiPRB}. The critical field $H_{c}$ for superconducting Eu is
thus quite large so that experiments to considerably higher magnetic fields
are required for its determination.

As for all trivalent rare-earth metals La through Lu, the conduction band of
the trivalent transition metals Sc and Y has $s,p,d$-electron character where
the $d$-electron concentration increases with pressure. It is thus not
surprising that under pressure Y fits nicely into the crystal-structure
sequence observed across the rare-earth series; Sc appears to follow a
different structure sequence, perhaps because it is much lighter
\cite{akahama}. We note that under exteme pressure the values of the
superconducting transition temperature for Sc, Y, La, and Lu all lie in the
range 10 - 20 K. One may thus ask why $T_{c}$ for Eu remains at temperatures
below 3 K, even at extreme pressures well over 1 Mbar. One possibility for
this result is that the crystal structure of Eu in this pressure range is less
favorable for superconductivity. A second possibility is that Eu at 142 GPa is
indeed completely trivalent, implying a 4f$^{6}$ configuration, but that the
quantum mechanical mixing between the nonmagnetic $J=0$ ground state and the
low-lying magnetic $J=1$ excited state, which is responsible for the Van Vleck
paramagnetism, weakens the superconducting pairing interaction and lowers
$T_{c}.$ A third possibility is that for $P\leq$ 142 GPa Eu metal is not fully
trivalent, but rather intermediate valent and that the fluctuations between
magnetic 4$f^{7}$ and nonmagnetic 4f$^{6}$ ground state configurations either
weaken the superconductivity or, alternatively, even help mediate it as has
been suggested for a number of Ce compounds \cite{yuan}.

In summary, Eu metal, a divalent antiferromagnet at ambient pressure, is found
to become superconducting near 1.8 K for pressures above 80 GPa, where $T_{c}$
increases with pressure at the very moderate rate of +18 mK/GPa to 2.75 K at
142 GPa. Whether the superconductivity occurs in a trivalent or mixed-valent
state of Eu is not yet clear. To more fully explore the fascinating interplay
of superconductivity, magnetism, and valence transition in Eu under pressure,
future measurements should examine both the pressure-dependent magnetic order
and superconductivity of Eu to multi-Mbar pressures, including the direct
determination of Eu's valence through L$_{\text{III}}$ absorption edge and/or
M\"{o}ssbauer-effect studies.

\vspace{0.5cm}\noindent Acknowledgments. Thanks are due to R. W. McCallum and
K. W. Dennis of the Materials Preparation Center, Ames Laboratory, for
providing the high-purity Eu sample. The authors thank V. K. Vohra for
recommending the specifications of the beveled diamond anvils used in these
experiments. Helpful suggestions from M. M. Abd-Elmeguid, A. K. Gangopadhyay,
B. Johansson, J. A. Mydosh, and F. Steglich are gratefully acknowledged. The
research visit of one of the authors (T. M.) at Washington University was made
possible by the Osaka University short-term student dispatch program. The
authors gratefully acknowledge research support by the National Science
Foundation through Grant No. DMR-0703896.

\begin{center}
\bigskip{\LARGE Figure Captions}
\end{center}

\bigskip\ 

\noindent\textbf{Fig. 1. \ }\bigskip(color online) Quasi four-point electrical
resistance measurements versus temperature for Eu metal at 37, 48, 61, 73, 81,
91, and 62 GPa, taken in that order. Inset shows data near 2 K for 73, 81, and
91 GPa on a highly expanded scale. $T_{c}$ is determined from the
superconducting onset as the temperature is increased (see arrows).

\noindent\textbf{Fig.\ 2.\ \ }\bigskip(color online) Real part of the ac
susceptibility versus temperature for Eu metal as pressure is increased from
76 to 142 GPa. The superconducting transition appears at 84 GPa and shifts
slowly under pressure to higher temperatures. $T_{c}$ is determined from the
temperature at the transition midpoint. The inset shows raw $\chi^{\prime}(T)$
data at 118 GPa (see Ref. \cite{note2}).

\noindent\textbf{Fig. 3. \ }(color online) Superconducting transition
temperature of Eu metal versus pressure from electrical resistivity ($\oplus$)
and ac susceptibility ($\blacksquare$) data in Figs.~1 and 2, respectively.
Vertical error bars\ give 20\% - 80\% transition width. Solid straight line is
guide to the eye.


\begin{thebibliography}{99}
\bibitem {note2}As discussed in Ref.\cite{DebessaiPRB}, a
temperature-dependent background signal has been subtracted from the
susceptibility data in Fig. 2. The inset shows the raw data at 118 GPa before
this background subtraction. The fact that superconductivity above 1.5 K
appears at a somewhat lower pressure in resistivity than in magnetic
susceptibiity measurements likely arises from uncertainties in the estimated
pressure at the sample due to pressure gradients and differences in the
criteria for determining the value of $T_{c}$.
\bibitem {note1}In all resistivity and ac susceptibility experiments on Eu, a
25 $\mu$m-thick Zr foil was placed under the table of the diamond anvil. To
verify directly that the observed diamagnetic jump $\Delta\chi^{\prime} $ near
2 K does not originate from superconductivity in the Zr foil, this foil was
removed in a test experiment. At 84 GPa the strong 20 nV diamagnetic
transition from Eu appeared, as usual, at $\sim$ 1.7 K, where none had been
observed above 1.5 K at 75 GPa; in this configuration the Eu sample was then
removed, whereupon the diamagnetic transition disappeared above 1.3 K to
pressures as high as 94 GPa.
\end{thebibliography}
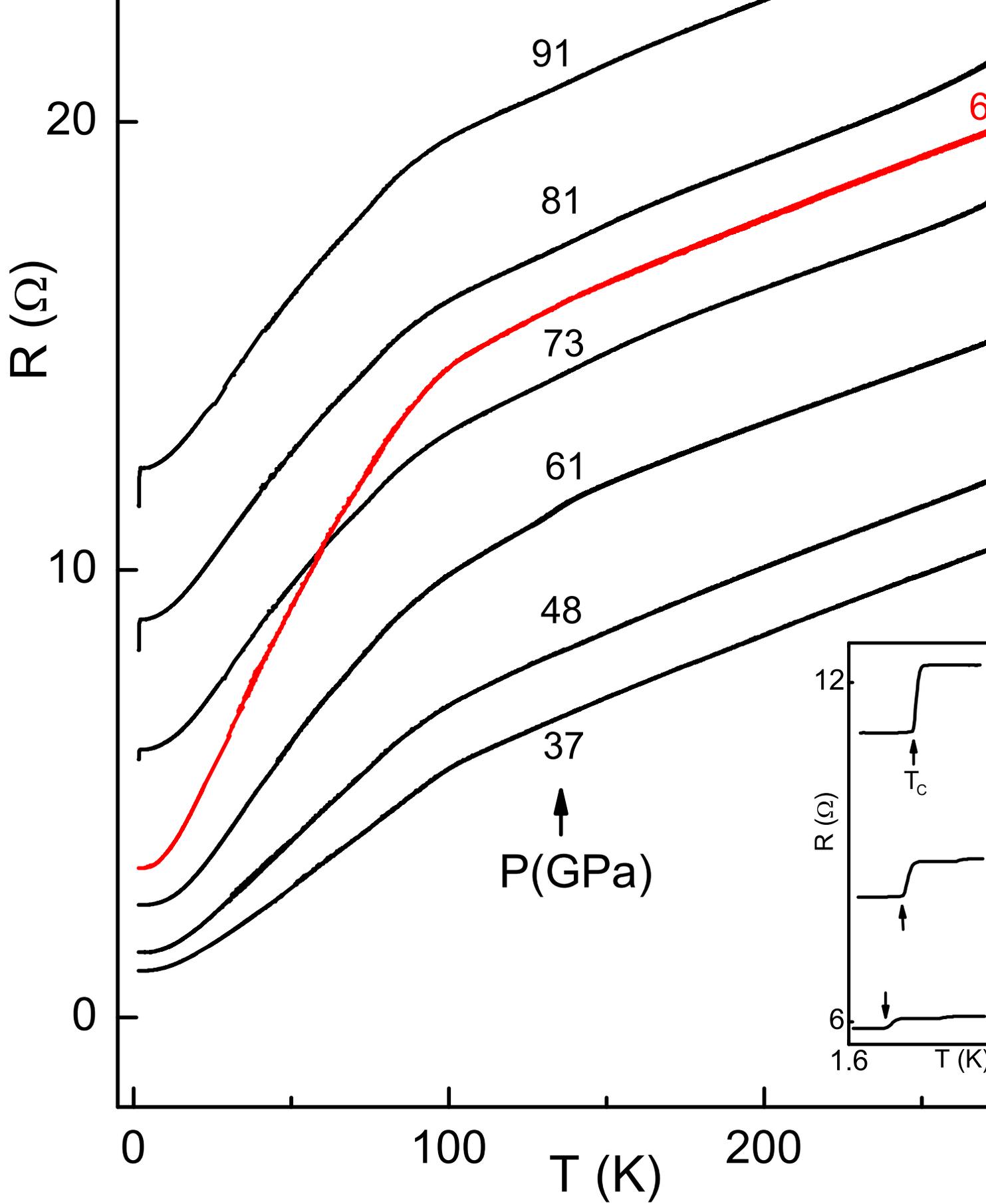
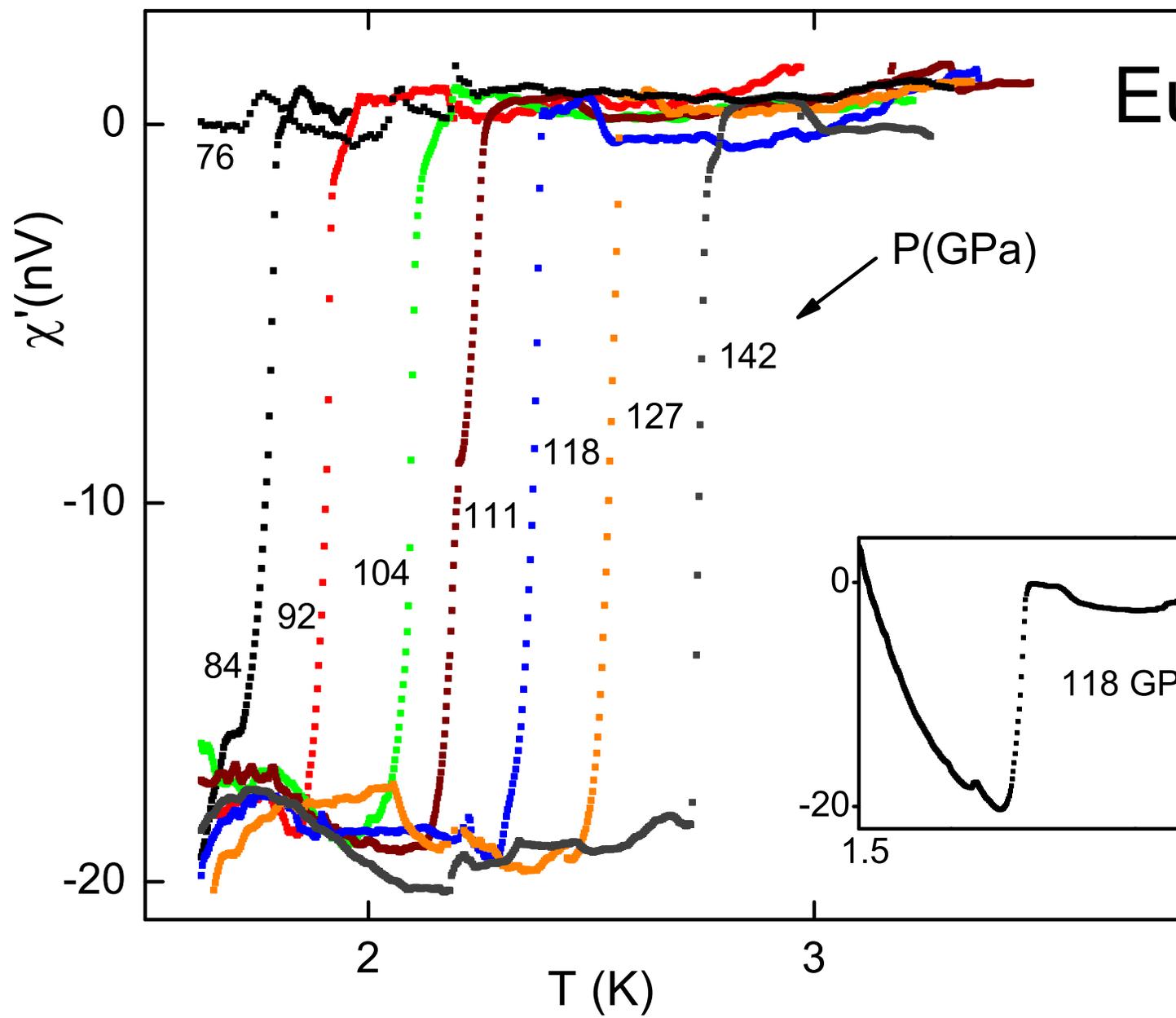
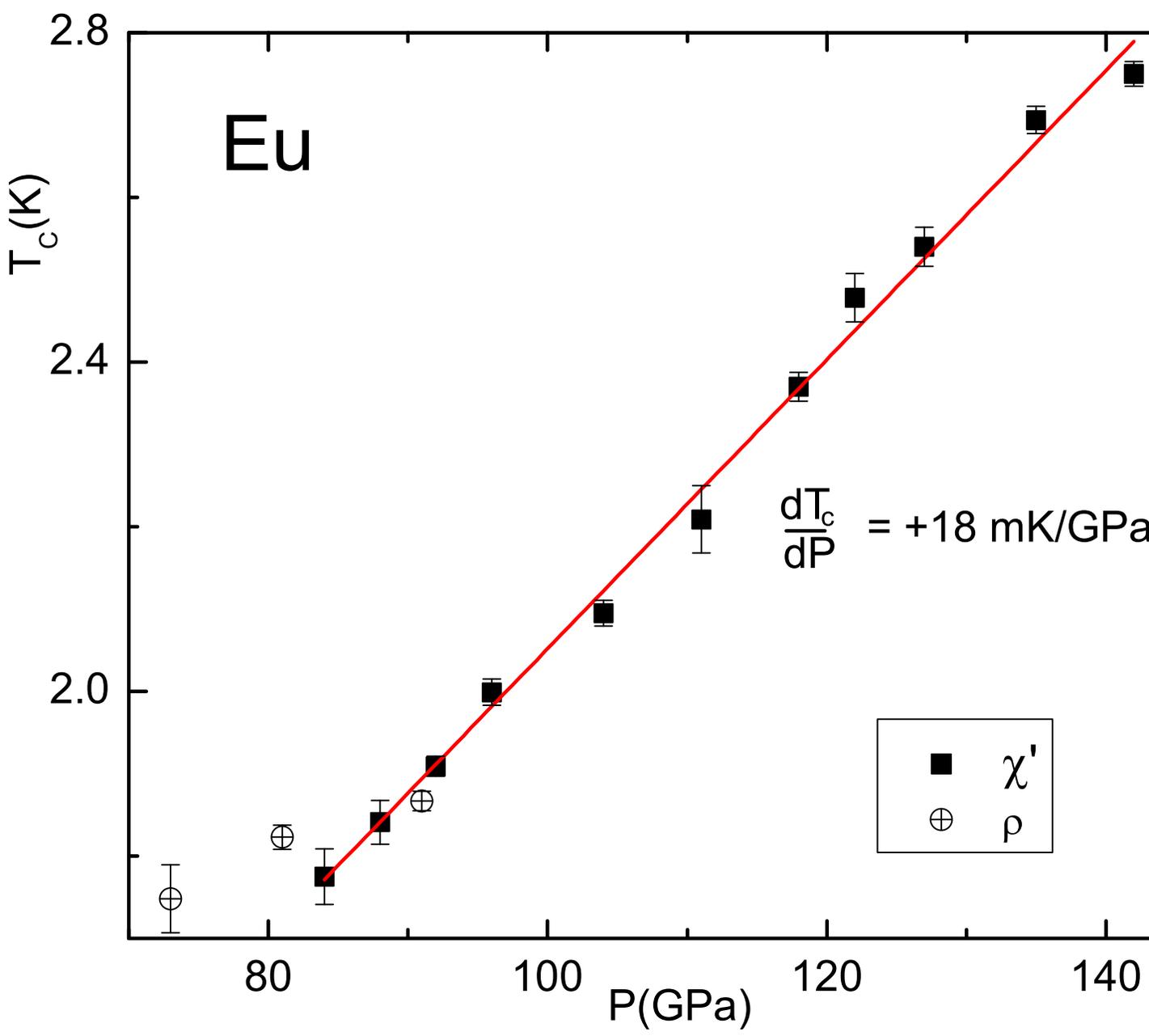
\end{document}